\documentclass[runningheads,a4paper]{llncs}

\usepackage{amssymb}
\usepackage{graphicx}
\usepackage{boxedminipage}

\usepackage{amsmath}
\usepackage{algorithm}
\usepackage{algorithmic}

\usepackage{url}
\urldef{\mailsa}\path|{alfred.hofmann, ursula.barth, ingrid.haas, frank.holzwarth,|
\urldef{\mailsb}\path|anna.kramer, leonie.kunz, christine.reiss, nicole.sator,|
\urldef{\mailsc}\path|erika.siebert-cole, peter.strasser, lncs}@springer.com|

\begin{document}

\mainmatter  

\title{Efficient Three-party Computation: \\ An Information-theoretic Approach from Cut-and-Choose}

\titlerunning{Lecture Notes in Computer Science: Authors' Instructions}

%
%
\author{Zhili Chen%
}
%
%
\institute{Anhui University,\\
Hefei, Anhui 230601, China\\
zlchen@ahu.edu.cn
}
%
%
\maketitle

\begin{abstract}
As far as we know, the literature on secure computation from cut-and-choose has focused on achieving computational security against malicious adversaries. It is unclear whether the idea of cut-and-choose can be adapted to secure computation with information-theoretic security.

In this work we explore the possibility of using cut-and-choose in information theoretic setting for secure three-party computation (3PC). Previous work on 3PC has mainly focus on the semi-honest case, and is motivated by the observation that real-word deployments of multi-party computation (MPC) seem to involve few parties. We propose a new protocol for information-theoretically secure 3PC tolerating one malicious party with cheating probability $2^{-s}$ using $s$ runs of circuit computation in the cut-and-choose paradigm. The computational cost of our protocol is essentially only a small constant worse than that of state-of-the-art 3PC protocols against a semi-honest corruption, while its communication round is greatly reduced compared to other maliciously secure 3PC protocols in information-theoretic setting.

\end{abstract}

\section{Introduction}

Secure multi-party computation (MPC) allows a set of parties to compute a  function of their joint inputs without revealing anything beyond the fuction output. During the past few years, a tremendous amount of attention has been devoted to making MPC practical, such as \cite{huang2011faster} \cite{keller2013architecture}. However, most of such works have focused on secure two-party computation (2PC). In the semi-honest setting, it has been shown that Yao's garbled circuit technique \cite{86yaoa} can yield very efficent protocols for the computation of boolean circuits \cite{ben2008fairplaymp,henecka2010tasty,huang2011faster,huang2012private}. In the malicious setting, the cut-and-choose technique \cite{lindell2007efficient} based on Yao's garbled circuit was used to construct efficient, constant-round protocols. This technique was further developed in a large body of subsequent work \cite{woodruff2007revisiting,lindell2008implementing,nielsen2009lego,shen2011two,lindell2012secure,kreuter2012billion,huang2013efficient} \cite{lindell2013fast,mohassel2013garbled,shen2013fast}, yielding the fastest protocols for 2PC of Boolean circuits.

Recently, Yao's garbled circuits are also applied to constructing secure three-party computation (3PC) against malicious corruptions, with \cite{choi2014efficient} and without \cite{mohassel2015fast} the cut-and-choose technique. It appears that the cut-and-choose technique is naturally used with Yao's garbled circuits. It is unknown so far if this idea can be applied to the information-theoretic setting.

Secure three-party computation (3PC) where only one party is corrupted (honest majority) is an important special case where the least number of parties is satisfied for achieving information-theoretic security. In this setting, protocols designed with semi-honest security can be significantly more efficient than their two-party counterparts since they are commonly based on secret sharing schemes instead of cryptographic operations. To achieve security against malicious adversaries, usually verifiable secret sharing is used, resulting significant communication complexity and overhead. Existing work includes implementation and optimization in frameworks such as Sharemind \cite{bogdanov2008sharemind}, ShareMonad \cite{launchbury2012efficient,launchbury2014application} .

\subsection{Our Contribution}

We design a new protocol for 3PC with one malicious corruption from cut-and-choose in the information-theoretic setting. Unlike the standard approach of applying cut-and-choose technique for compiling GC-based protocols into malicious 2PC, we show that in the setting of information-theoretically secure 3PC with one corruption one can also use the cut-and-choose paradigm and achieve malicious security at a cost similar to semi-honest 3PC constructions.

The communication complexity of our protocol is linear to the depth of the circuit, just as that of the common methods of information theoretically secure computations (e.g., the verifiable secret sharing methods). However, through the application of cut-and-choose, we can significantly reduce the communication rounds greatly, comparied with the normal methods from verifiable secret sharing.

In practical applications, our protocol can be converted into one against convert adversaries \cite{lindell2013fast} by fixing the statsitcal parameter $s$ at a small value, yeilding a faster protocol. For example, if the covert securlty level, where the protocol ensures that $99.5\%$ of cheats are caught, is sufficient for practical application, then fixing $s = 8$ is good, and the protocol is secure in the presence of convert adversaries with deterrent $\epsilon = 1 - 2^{-8}$.

\subsection{Overview of Our Protocol}
The high-level idea of our construction is to execute the three-party protocol against semi-honest adversaries $s$ times in parallel. The transcripts of $r$ randomly selected runs are revealed for verifying the correctness of computation (i.e., \emph{computation verification}), while the remaining $s-r$ runs are used for computing the output (i.e., \emph{output computation}). This is a classical \emph{cut-and-choose paradigm}. However, to achieve malicious security, there are two challenges as follows. First, the randomly selected runs for \emph{computation verification} should not use the true input of each party, while those for \emph{output computation} should. Otherwise, the privacy and/or correctness of the protocol would be violated. Meanwhile, the selection randomness (for using true inputs or not) should be unknown to any participant party, but can be determined cooperatively by all the three parties. The reason is that if any party knows this randomness, it can easily make modification to the \emph{output computation} runs without being caught in the cut-and-choose process.

\begin{figure}
\centering
\includegraphics[width=0.8\textwidth]{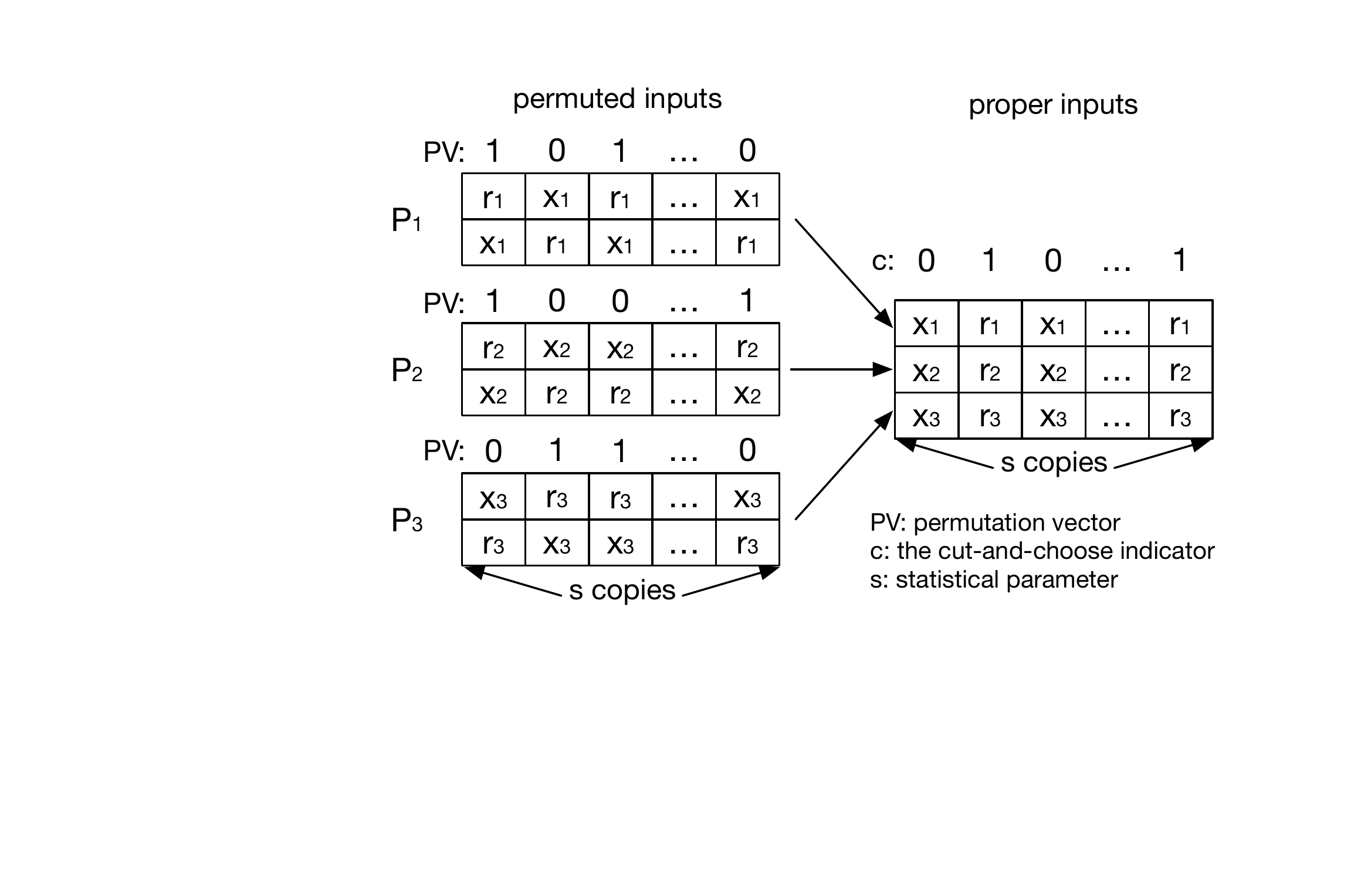}
\caption{Each party generates $s$ copies of permuted inputs with a privately generated permutation vector. The proper inputs are then computed from these permuted inputs according to a cut-and-choose indicator.} \label{fig:overview}
\end{figure}

To address the first challenge, in our protocol, we let each party prepare two versions of input: one is the \emph{true input} and the other is a \emph{random input} (i.e., a randomly selected string of the same bit length as the true input). Next, let each party make $s$ copies of its input, with its two versions of input randomly permuted in each copy. A well-designed circuit is then used to select a proper input according to a cut-and-choose indicator $c$ for each run. Specifically, if the $j$th bit of $c$ is $1$, then the true input is selected; otherwise, the random input is selected (see Fig. \ref{fig:overview}).

In order to address the second challenge, the permutation randomness of the two input versions of each party is privately generated, and unknown to any other party. Thus, the shared inputs cannot be altered, without being detected, by any party except its owner. This prevents the malicious party from tampering the inputs without being caught. Additionally, the cut-and-choose indicator $c$ is generated by the three parties in a way such that,  no individual party knows anything about $c$, but all of them can determine $c$ cooperatively. This ensures that any party cannot determine whether a given run is used for computation verification or output computation and thus cannot falsify the circuit computation without being caught.

\subsection{Construction}
In Section \ref{sec:prelim}, some preliminary topics including the circuit notation, secret sharing, security model are described. In Section \ref{sec:3pc-semi}, the underlying 3PC against semi-honest adversaries is introduced. Our maliciously secure 3PC from cut-and-choose is detailed and analyzed in Section  \ref{sec:3pcfromcc}. Then, some practical considerations of our protocol is mensioned in Section \ref{sec:prac-consid}. Finally, Section \ref{sec:conclusion} is the conclusion of our work.

\section{Preliminaries}\label{sec:prelim}

We let $s$ denote the statistical security parameter, and assume that the computed function is represented exclusively with XOR and AND gates.

\subsection{Circuit Notation}

We follow the circuit notation in \cite{bellare2012foundations}. Denote a circuit $C = (n, m, q, F, S, G)$, where $n$ and $m$ are the numbers of input and output wires, respectively, $q$ is the number of gates, and each gate is indexed by its output wire. Obviously, the total number of wires in the circuit is $n + q$. Let the wire numbering start with input wires and end with output wires, then the input wires are $\{1,\cdots,n\}$ and the output wires are $\{n + q - m + 1, \cdots, n + q\}$. The function $F$ (resp., $S$) takes as input a gate index and returns the first (resp., second) input wire to gate, with a constraint $F(g) < S(g) < g$ for any gate index $g$. The function $G$ represents the operation of a given gate, e.g., $G_g(0, 1) = 1$ if the gate with index $g$ is an XOR gate. Since we consider circuits with inputs from three parties, let $\{n_{i-1} + 1, \cdots, n_i\}$ denote the input wires of party $P_i$ for $i \in \{1,2,3\}$, with $n_0 = 0$ and $n_3 = n$.

\subsection{XOR Secret Sharing}

Our constructions use three-out-of-three XOR secret sharing: the secret $x \in \{0,1\}^k$ is split into three random shares $x_1$, $x_2$ and $x_3 \in \{0, 1\}^k$ such that $x_1 \oplus x_2 \oplus x_3 = x$, with $P_i$ holding share $x_i$ for $i \in \{1,2,3\}$. We denote a share of $x$ by $[x]_i = x_i$ and the sharing by $[x] = ([x]_1, [x]_2, [x]_3)$. This sharing is linear: if $[x]$ and $[y]$ are sharings of $x$ and $y$ respectively, then $[x] \oplus [y]$ is a sharing of $x \oplus y$, i.e., $[x \oplus y] = [x] \oplus [y]$. This means, each party $P_i$ ($i \in \{1, 2, 3\}$) can locally compute his share of $x \oplus y$ as $[x \oplus y]_i = [x]_i \oplus [y]_i$.

It is straight-forward to show that the above secret sharing scheme is information theoretically secure given that the shares $x_1$, $x_2$ and $x_3$ are uniformly chosen (subject to $x_1 \oplus x_2 \oplus x_3 = x$). Reconstructing a sharing $[x]$ is easily done by having each party $P_i$ announce his share $[x]_i$ and taking $x = [x]_1 \oplus [x]_2 \oplus [x]_3$.

\subsection{Shamir's Secret Sharing}

Shamir's secret sharing \cite{shamir1979share} is a fundamental $(t, n)$-threshold secret sharing scheme. The related notions are described as follows.

\textbf{$(t,n)$-threshold secret sharing.} A secret $s$ is dispersed into $n$ shares in such a way that any
$k \le t$ shares give no information on $s$ (i.e., $t$-privacy), whereas any $k \ge t + 1$ shares uniquely determine $s$ (i.e., $(t + 1)$-reconstruction), where $t$,
$n$ are integers with $0 \le t < n$.

In $(t,n)$-threshold secret sharing, if an adversary obtains at most $t$ shares, he obtains nothing about the secret. One of the widely used
$(t,n)$-threshold secret sharing schemes is Shamir's defined as follows.

\textbf{Shamir's $(t,n)$-threshold Secret Sharing. } Let $p > n$ be a prime. Given integers $t$, $n$ with $0 \le t <
n$, and $\alpha_1, \alpha_2, ..., \alpha_n \in \mathbb{F}_p$ being
pairwise distinct and non-zero. Note that $p$, $t$, $n$, $\alpha_1,
\alpha_2, ..., \alpha_n$ are public data. Denote the secret by $s \in
\mathbb{F}_p$. Select a polynomial $f(x) \in
\mathbb{F}_p[X]$ uniformly at random, conditioned on $deg(f) \le t$
and $f(0) = s$. The $n$ shares in the secret $s$ are then given as
follows:
\[\small
s_i = f(\alpha_i) \in \mathbb{F}_p, \text{ for } 1 \le i \le n.
\]

It is proved that Shamir's secret sharing satisfies $t$-privacy and $(t+1)$-reconstruction
with $t + 1 \le n$.

\textbf{Reconstruction of Shamir's Secret Sharing}. Given a set of shares corresonding to $C \subset \{\alpha_i| 1 \le i <n\}$ with $|C| = t + 1$, the reconstruction of $s$ is as follows.

First, the polynomial $f(x)$ over $\mathbb{F}$ is reconstructed using Lagrange interpolation.
\begin{equation}
f(x) = \sum_{i \in C} f(\alpha_i) \theta_i(x)
\end{equation}
where $\theta_i(x)$ is the degree $t$ polynomial such that, for all $i, j \in C$, $\theta_i(\alpha_j) = 0$ if $i \neq j$ and $\theta_i(\alpha_j) = 1$ if $i = j$, namely,
\begin{equation}
\theta_i(x) = \prod_{\alpha \in C, j \neq i} \frac{x - \alpha_j}{\alpha_i - \alpha_j}
\end{equation}

Then, $s$ is the output of function $f(x)$ at $x = 0$, namely $s = f(0)$.

\subsection{Universally Composable (UC) Security}

The framework of UC security involves a collection of interactive Turing machines, and security is defined by comparing a real and ideal interaction among these machines. In the real interaction, parties cooperatively run a protocol, while an adversary may corrupt a part of parties. In the ideal interaction, parties run a dummy protocol by sending inputs to and receiving outputs from an uncorruptable \emph{functionality} machine, which performs the entire computation on behalf of the parties. An \emph{environment} machine is introduced to both real and ideal interactions, representing anything external to the current protocol execution. The \emph{environment} provides all inputs to all parties and reads all outputs from them, and can also interact arbitrarily with the \emph{adversary} (resp. \emph{simulator}) in the real (resp. ideal) interaction. Security means that the environment cannot distinguish the real and ideal interactions. At the end of interaction, the \emph{environment} output a bit.

Let $\mathtt{REAL}[\mathcal{Z}, \mathcal{A}, \pi]$ and $\mathtt{IDEAL}[\mathcal{Z}, \mathcal{S}, \mathcal{F}]$ denote the output of the \emph{environment} $\mathcal{Z}$ when interacting with \emph{adversary} $\mathcal{A}$ and parties who execute protocol $\pi$, and when interacting with \emph{simulator} $\mathcal{S}$ and parties run the dummy protocol in the presence of \emph{functionality} $\mathcal{F}$, respectively. We say that protocol $\pi$ is \textbf{information-theoretically secure} if for every \emph{adversary} $\mathcal{A}$ with unbounded computation power attacking the \emph{real} interaction, there exists a \emph{simulator} $\mathcal{S}$ attacking the \emph{ideal} interaction, such that for all \emph{environments} $\mathcal{Z}$, Eq. \eqref{eq:itsecurity} holds with negligible exception.
\begin{equation}\label{eq:itsecurity}
|Pr[\mathtt{REAL}[\mathcal{Z}, \mathcal{A}, \pi] = 1] - Pr[\mathtt{IDEAL}[\mathcal{Z}, \mathcal{S}, \mathcal{F}] = 1]| = 0
\end{equation}

Particularly, we say that \textbf{perfect security} is achieved, or the real view is \textbf{perfectly simulated} if Eq. \eqref{eq:itsecurity} holds.

\section{3PC in Semi-honest Setting}\label{sec:3pc-semi}

We use as a basis a 3PC protocol against semi-honest adversaries, a variant of ITSEC proposed in \cite{15chenz}. Let $h(i) = (i - 1) \% 3 + 1$, we have $h(i) = i$ and $h(i + 3) = h(i) = i$, for $i \in \{1,2,3\}$. Denote the bit value of wire $w$ by $\delta(w)$, and the $k$-bit string value of wire $w$ by $\delta^k(w)$. The detailed protocol of the variant ITSEC is illustrated in Fig. \ref{fig:itsec-variant}.

\begin{figure}
\centering
\begin{boxedminipage}{\linewidth}
\begin{center}\textbf{Protocol }$\pi_{\mathtt{3PC}}^{\mathtt{sem}}(P_1, P_2, P_3)$\end{center}
\textbf{Input:} Party $P_i$ holds input $x_i \in \{0,1\}^{n_i - n_{i-1}}$ with $i \in \{1,2,3\}$ \\
\textbf{Auxiliary Inputs:} Circuit $C = (n, m, q, F, S, G)$ containing exclusively XOR and AND gates
\begin{itemize}
\item[1. ]\textbf{Input sharing:}
\begin{itemize}
\item[ - ] For $i \in \{1,2,3\}$:

\begin{itemize}
\item[ $\diamond$ ] $P_i$ generates its input sharing $[x_i]$
\item[ $\diamond$ ] $P_i$ sends $[x_i]_{h(i+1)}$ to $P_{h(i+1)}$, and $[x_i]_{h(i+2)}$ to $P_{h(i+2)}$
\item[ $\diamond$ ] $P_i$ receives $[x_{h(i-1)}]_i$ from $P_{h(i-1)}$, and $[x_{h(i-2)}]_i$ from $P_{h(i-2)}$
\end{itemize}
\item[ - ] $P_i$ holds $([x_1]_i, [x_2]_i, [x_3]_i)$ for $i \in \{1,2,3\}$
\end{itemize}

\item[2. ]\textbf{Circuit computation:}\\
For $g \in \{n + 1, \cdots, n + q\}$: $a \leftarrow F(g)$, $b \leftarrow S(g)$
\begin{itemize}
\item[ - ] If $g$ is XOR gate: For $i \in \{1,2,3\}$: $[\delta(g)]_i \leftarrow [\delta(a)]_i \oplus [\delta(b)]_i$
\item[ - ] If $g$ is AND gate (see Fig. \ref{fig:and-com}): For $i \in \{1,2,3\}$:
\begin{itemize}
\item[ $\diamond$ ] $P_i$ generates a random bit $r_i$
\item[ $\diamond$ ] $P_i$ sends $[\delta(a)]_i$, $[\delta(b)]_i$ and $r_i$ to $P_{h(i+1)}$
\item[ $\diamond$ ] $P_i$ receives $[\delta(a)]_{h(i-1)}$, $[\delta(b)]_{h(i-1)}$ and $r_{h(i-1)}$ from $P_{h(i-1)}$
\item[ $\diamond$ ] $P_i$ computes \\
$[\delta(g)]_i \leftarrow [\delta(a)]_i [\delta(b)]_i \oplus [\delta(a)]_i [\delta(b)]_{h(i-1)} \oplus [\delta(a)]_{h(i-1)} [\delta(b)]_i \oplus r_i \oplus r_{h(i-1)}$
\end{itemize}
\end{itemize}

\item[3. ]\textbf{Output construction:}
\begin{itemize}
\item[ - ] For $g \in \{n + q - m + 1, \cdots, n + q\}$: For $i \in \{1,2,3\}$:
\begin{itemize}
\item[ $\diamond$ ] $P_i$ sends $[\delta(g)]_i$ to $P_{h(i+1)}$ and $P_{h(i+2)}$
\item[ $\diamond$ ] $P_i$ receives $[\delta(g)]_{h(i-2)}$ from $P_{h(i-2)}$ and $[\delta(g)]_{h(i-1)}$ from $P_{h(i-1)}$
\item[ $\diamond$ ] $P_i$ computes $\delta(g) \leftarrow [\delta(g)]_1 \oplus [\delta(g)]_2 \oplus [\delta(g)]_3$
\end{itemize}
\end{itemize}
\end{itemize}
\end{boxedminipage}
\caption{ITSEC Variant 3PC Protocol}
\label{fig:itsec-variant}
\end{figure}

\begin{figure}\label{fig:and-com}
\centering
\includegraphics[width=0.8\textwidth]{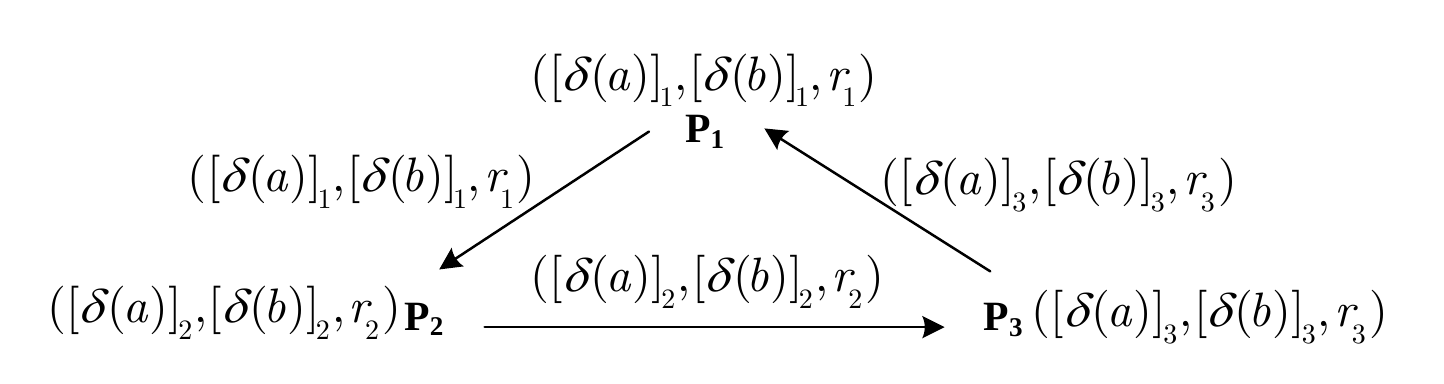}
\caption{The Communication Pattern for AND Gate Computation: $[\delta(a)\delta(b)]_i = [\delta(a)]_i [\delta(b)]_i \oplus [\delta(a)]_i [\delta(b)]_{h(i-1)} \oplus [\delta(a)]_{h(i-1)} [\delta(b)]_i \oplus r_i \oplus r_{h(i-1)}$ for $i \in \{1,2,3\}$. Note that this communication pattern uses no broadcast channels.}
\end{figure}

Theorem 1 states the security of the 3PC protocol $\pi_{\mathtt{3PC}}^{\mathtt{sem}}(P_1, P_2, P_3)$ in the semi-honest setting.\\\\
\textbf{Theorem 1. } \emph{Protocol $\pi_{\mathtt{3PC}}^{\mathtt{sem}}(P_1, P_2, P_3)$ is information-theoretically secure in the presence of a semi-honest adversary corrupting one party.}\\\\
\textbf{Proof.}  Since in protocol $\pi_{\mathtt{3PC}}^{\mathtt{sem}}(P_1, P_2, P_3)$, the three parties are symmetric,
we only need to prove the case where one of the parties is
corrupted.

Specifically, we show that for every adversary with unbounded
computation power, the environment's real view based on the interaction with the adversary and the parties executing the protocol is indistinguishable to the environment's simulated view based on the  interaction with a simulator and the the parties calling the functionality.

Without loss of generality, we assume that party $P_1$ is corrupted. For each phase of the protocol, we simulate the protocol execution as follows:
\begin{itemize}
\item \textbf{Input sharing}: Party $P_1$ receives $[x_3]_1$ from $P_3$, and $[x_2]_1$ from $P_2$. The two messages can be simulated by two uniformly random bit strings $s_1 \in \{0,1\}^{|x_3|}$ and $s_2 \in \{0,1\}^{|x_2|}$, resulting that both real and simulated views are identical.
\item \textbf{Circuit computation}:  This phase can be simulated gate by gate until the circuit computation is completed. Given a current gate, it is simulated in terms of its operation as follows.
\begin{itemize}
\item \textit{XOR}: Party $P_1$ sends or receives nothing, so there
is nothing to be simulated.
\item \textit{AND}: Let $a$ and $b$ denote the two input bits to the gate. During the gate computation, $P_1$ receives $[a]_3$,  $[b]_3$ and $r_3$ from $P_3$. From the circuit computation of the protocol, it is clear that bit shares $\{[a]_k\}_{k=1}^3$ and $\{[b]_k\}_{k=1}^3$ are independent uniformly random bits merely subject to $a = \oplus_{k=1}^3 [a]_k$ and $b = \oplus_{k=1}^3 [b]_k$, and $r_3$ is a uniformly generated random bit. Thus, the three messages can again be simulated by three uniformly random bits. Both the real and simulated views are also identical.
\end{itemize}

\item \textbf{Output Construction}: Let $y \in \{0,1\}^m$ be the output of the computed function. $P_1$ receives $[y]_3$ from $P_3$, and $[y]_2$ from $P_2$. Again, these two messages are uniformly random bit strings subject to $y = \oplus_{k=1}^3 [y]_k$, and can be simulated by two uniformly random bit strings $t_1, t_2 \in \{0,1\}^m$.

\end{itemize}
In all phases, we can see that both real and simulated views are identical. We thus conclude that Protocol $\pi_{\mathtt{3PC}}^{\mathtt{sem}}$ is information-theoretically secure against semi-honest adversaries corrupting one party. $\Box$

\section{Three-Party Computation from Cut-and-Choose} \label{sec:3pcfromcc}

In this section, we compile the 3PC protocol against semi-honest adversaries introduced in the previous section into a maliciously secure protocol using the cut-and-choose paradigm. This is the first time of applying cut-and-choose idea to the information-theoretic setting as far as we know. Then, the security and efficiency of the proposed protocol is analyzed.

\subsection{The Detailed Protocol}

Our 3PC protocol with malicious security can be described in four phases, each of which is detailed as follows.

\textbf{Input preparation.} Each party $P_i$ ($i \in \{1,2,3\}$) prepares a random input and its true input. That is, $P_i$ selects a random bit string $x^1_i$ (i.e., random input) of exactly the same bit length of its true input $x^0_i \in \{0,1\}^{n_i - n_{i-1}}$. Using $x^0_i$ and $x^1_i$, party $P_i$ generates $s$ copies of inputs.  We denote $P_i$'s $j$th copy of input by a 2-tuple $I_{ij} = (x^{\sigma_{ij}}_i, x^{1 - \sigma_{ij}}_i)$ with permutation bit $\sigma_{ij} \in \{0,1\}$. In the copy $I_{ij}$, if $\sigma_{ij} = 0$, the true input $x^0_i$ is placed in the first field; otherwise, it is placed in the second field. Let $I_i = (I_{i,1}, \cdots, I_{i,s})$ denote $P_i$'s set of inputs, with a random permutation vector $\sigma_i = (\sigma_{i,1}, \cdots, \sigma_{i,s})$ chosen by $P_i$. Note that this permutation vector is only know to the owner of the inputs, but unknown to other parties. This prevents other parties from modifying the true inputs but leaving the random inputs untouched.

Then, $P_i$ prepares the sharing of his input $[I_i] = ([I_{i,1}], \cdots, [I_{i,s}])$, where $[I_{i,j}] = ([x^{\sigma_{i,j}}_i], [x^{1 - \sigma_{i,j}}_i])$, and sends two of the shares to the other two parties, one for each party. After that, $P_i$ selects a random bit string $c_i = \{0,1\}^s$ as a share of the cut-and-choose indicator $c$, i.e., $[c]_i = c_i$, where the $j$th bit of $c$ indicates whether the $j$th run uses true inputs (bit 0) or random inputs (bit 1). Finally, $P_i$ commits to $x^1_i$ and $[c]_i$. It is worth noting here that no party knows the value of $c$ unless the commitments to all shares of $c$ is opened.

\begin{figure}
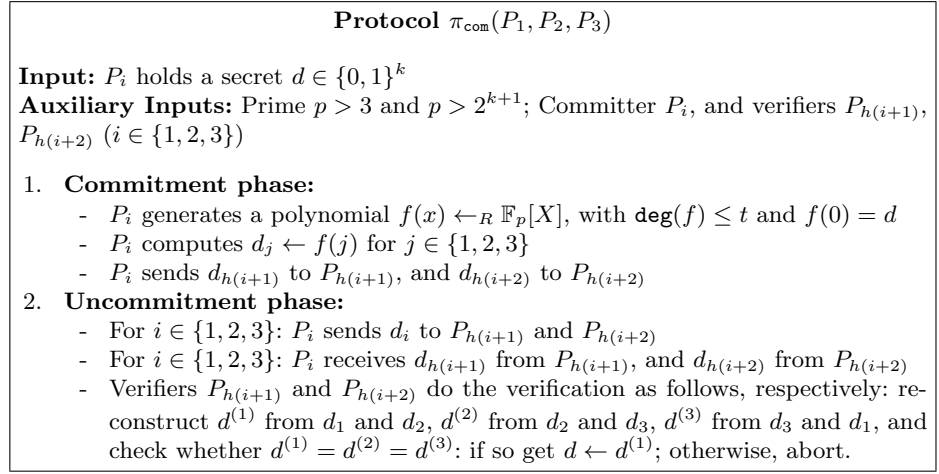

\centering
\begin{boxedminipage}{\linewidth}
\begin{center}\textbf{Protocol }$\pi_{\mathtt{com}}(P_1, P_2, P_3)$\end{center}
\textbf{Input:} $P_i$ holds a secret $d \in \{0,1\}^k$ \\
\textbf{Auxiliary Inputs:} Prime $p > 3$ and $p > 2^{k+1}$; Committer $P_i$, and verifiers $P_{h(i+1)}$, $P_{h(i+2)}$ ($i \in \{1,2,3\}$)
\begin{itemize}
\item[1. ]\textbf{Commitment phase:}
\begin{itemize}
\item[ - ] $P_i$ generates a polynomial $f(x) \leftarrow_R \mathbb{F}_p[X]$, with $\mathtt{deg}(f) \le t$ and $f(0) = d$
\item[ - ] $P_i$ computes $d_j \leftarrow f(j)$ for $j \in \{1,2,3\}$
\item[ - ] $P_i$ sends $d_{h(i+1)}$ to $P_{h(i+1)}$, and $d_{h(i+2)}$ to $P_{h(i+2)}$
\end{itemize}

\item[2. ]\textbf{Uncommitment phase:}
\begin{itemize}
\item[ - ] For $i \in \{1,2,3\}$: $P_i$ sends $d_i$ to $P_{h(i+1)}$ and $P_{h(i+2)}$
\item[ - ] For $i \in \{1,2,3\}$: $P_i$ receives $d_{h(i+1)}$ from $P_{h(i+1)}$, and $d_{h(i+2)}$ from $P_{h(i+2)}$

\item[ - ] Verifiers $P_{h(i+1)}$ and $P_{h(i+2)}$ do the verification as follows, respectively: reconstruct $d^{(1)}$ from $d_1$ and $d_2$, $d^{(2)}$ from $d_2$ and $d_3$, $d^{(3)}$ from $d_3$ and $d_1$, and check whether $d^{(1)} = d^{(2)} = d^{(3)}$: if so get $d \leftarrow d^{(1)}$; otherwise, abort.
\end{itemize}

\end{itemize}
\end{boxedminipage}
\caption{Commitment Scheme based on Shamir's Secret Sharing}
\label{fig:commit}
\end{figure}

For commitment, we use scheme $\pi_{\mathtt{com}}(P_1, P_2, P_3)$ depicted in Fig. \ref{fig:commit} in the whole paper unless noted otherwise. This commitment scheme is a three-party protocol and based on Shamir's secret sharing. It is of perfect hiding and perfect binding as we will show in Section \ref{sec:analysis}. Furthermore, the scheme has other properties as follows. First, it is uncommitment-symmetric: in the commitment phase, one party (the committer) commits to the value, while in the uncommitment phase, any party can uncommit the committed value. Second, it is addition (substraction) homomorphic. That is, given commitments to $a$ and $b$, it is easy for each party to locally compute the commitments to $a+b$ and $a-b$ by adding or substracting the corresponding shares. These two properties are straight-forward to verify.

\begin{figure}
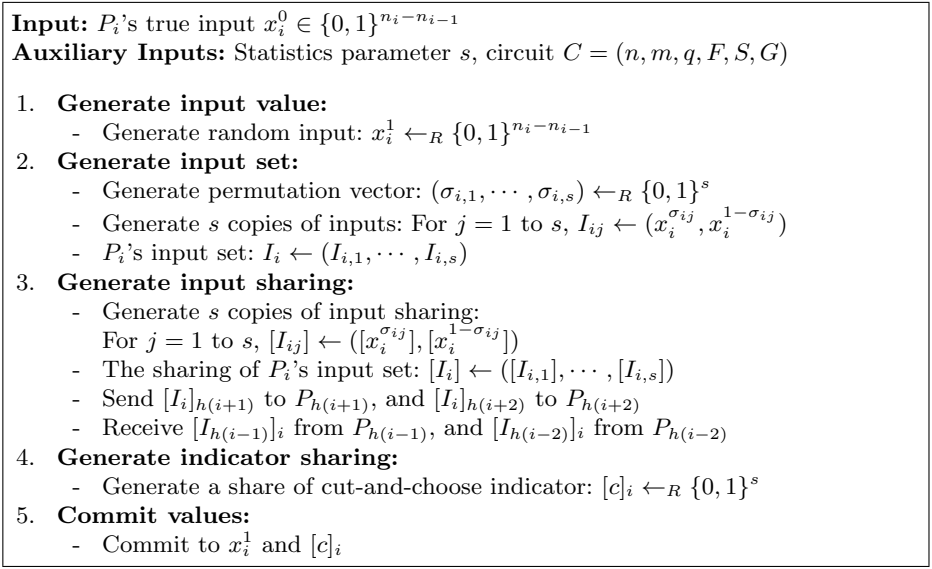

\centering
\begin{boxedminipage}{\linewidth}
\textbf{Input:} $P_i$'s true input $x^0_i \in \{0,1\}^{n_i - n_{i-1}}$ \\
\textbf{Auxiliary Inputs:} Statistics parameter $s$, circuit $C = (n, m, q, F, S, G)$
\begin{itemize}
\item[1. ]\textbf{Generate input value:}
\begin{itemize}
\item[ - ] Generate random input: $x^1_i \leftarrow_R \{0,1\}^{n_i - n_{i-1}}$
\end{itemize}

\item[2. ]\textbf{Generate input set:}
\begin{itemize}
\item[ - ] Generate permutation vector: $(\sigma_{i,1}, \cdots, \sigma_{i,s}) \leftarrow_R \{0, 1\}^s$
\item[ - ] Generate $s$ copies of inputs: For $j = 1$ to $s$, $I_{ij} \leftarrow (x^{\sigma_{ij}}_i, x^{1 - \sigma_{ij}}_i)$
\item[ - ] $P_i$'s input set: $I_i \leftarrow (I_{i,1}, \cdots, I_{i,s})$
\end{itemize}

\item[3. ]\textbf{Generate input sharing:}
\begin{itemize}
\item[ - ] Generate $s$ copies of input sharing: \\ For $j = 1$ to $s$, $[I_{ij}] \leftarrow ([x^{\sigma_{ij}}_i], [x^{1 - \sigma_{ij}}_i])$
\item[ - ] The sharing of $P_i$'s input set: $[I_i] \leftarrow ([I_{i,1}], \cdots, [I_{i,s}])$
\item[ - ] Send $[I_i]_{h(i+1)}$ to $P_{h(i+1)}$, and $[I_i]_{h(i+2)}$ to $P_{h(i+2)}$
\item[ - ] Receive $[I_{h(i-1)}]_i$ from $P_{h(i-1)}$, and $[I_{h(i-2)}]_i$ from $P_{h(i-2)}$
\end{itemize}

\item[4. ]\textbf{Generate indicator sharing:}
\begin{itemize}
\item[ - ] Generate a share of cut-and-choose indicator: $[c]_i \leftarrow_R \{0, 1\}^s$
\end{itemize}

\item[5. ]\textbf{Commit values:}
\begin{itemize}
\item[ - ] Commit to $x^1_i$ and $[c]_i$
\end{itemize}

\end{itemize}
\end{boxedminipage}
\caption{Input Preparation of $P_i$ ($i \in \{1,2,3\}$)}
\label{fig:input-preparation}
\end{figure}

The detailed protocol of input preparation is shown in Fig. \ref{fig:input-preparation}. Then after every party executes its input preparation, $P_i$ holds $([I_1]_i, [I_2]_i, [I_3]_i)$, and the commitments to $x^1_i$ and $[c]_i$.

\textbf{Circuit computation.} The three parties cooperatively compute the function $s$ times in parallel, using the prepared input sharings $[I_i]$ ($i \in \{1,2,3\}$) and the sharing $[c]$ of cut-and-choose indicator. Unlike the semi-honest setting, the cut-and-choose indicator should be used in the malicious scenario to select either the true or random inputs of the three parties, such that if the bit of $c$ is 1, the inputs to the corresponding run are random inputs; otherwise, they are the true inputs. Thus, we have to design a circuit (called input selection circuit) for each run to select proper inputs using $\sigma_{ij}$, $I_{ij}$ and $c$ with $i \in \{1,2,3\}$ and $j \in \{1,\cdots,s\}$.

Since for each $\sigma_{ij}$, $I_{ij} = (x^{\sigma_{ij}}_i, x^{1-\sigma_{ij}}_i)$, we have
\begin{align}
x^0_i &= (x^{\sigma_{ij}}_i \oplus x^{1-\sigma_{ij}}_i)\sigma_{ij} \oplus x^{\sigma_{ij}}_i \label{x0i} \\
x^1_i &= (x^{\sigma_{ij}}_i \oplus x^{1-\sigma_{ij}}_i)\sigma_{ij} \oplus x^{1 - \sigma_{ij}}_i \label{x1i}
\end{align}

Let $c = (c^{(1)}, \cdots, c^{(s)})$, where $c^{(j)}$ is the $j$th bit of $c$. For $j$th run, the input selection circuit can be designed as follows.
\begin{align}
x_{ij} = (x^0_i \oplus x^1_i)c^{(j)} \oplus x^0_i \label{xij}
\end{align}
where $x_{ij}$ is the input of $P_i$ picked up using the cut-and-choose indicator in the $j$th run (i.e., if $c^{(j)} = 0$ then $x_{ij} = x^0_i$; otherwise, $x_{ij} = x^1_i$).

Plug Eqs. \eqref{x0i} and \eqref{x1i} into Eq. \eqref{xij}, we then get
\begin{align}
x_{ij} &= (x^{\sigma_{ij}}_i \oplus x^{1 - \sigma_{ij}}_i)c^{(j)} \oplus (x^{\sigma_{ij}}_i \oplus x^{1-\sigma_{ij}}_i)\sigma_{ij} \oplus x^{\sigma_{ij}}_i \label{xij2}\\
       &= (x^{\sigma_{ij}}_i \oplus x^{1 - \sigma_{ij}}_i)(c^{(j)} \oplus \sigma_{ij}) \oplus x^{\sigma_{ij}}_i \label{xij3}
\end{align}
Concatenating the circuit represented by Eq. \eqref{xij3} to the circuit of the computed function, the three parties can cooperatively decide the inputs (true or random) to each run, without any party knowing anything about the decision.

Then, with each shared and decided copy of input, the three parties cooperatively compute the circuit of the computed function once. Each party records the transcripts (values of all the wires) of all the $s$ runs.

\textbf{Transcript commitment.} Each party $P_k$ commits to the transcript of every run of the circuit. The transcript of a circuit $C = (n, m, q, F, S, G)$ can be described with all its wire states by a string $t \in \{0,1\}^{n+q}$.  We let $P_k$ stores its transcript in an array $T^j_k[1..n+q]$ for its $j$th run.  Then, $T^j_k[1..n]$ denotes the input share of $P_k$,  $T^j_k[n + 1..n + q - m]$ denotes the share of internal gate outputs (called internal state share), and $T^j_k[n + q - m + 1..n + q]$ denotes the output share. Each party commit to its input share, interal state share and output share for every run, respectively. That is, Each party commits to its share on every wire in very run. These commitments make all parties cannot change their data shares in the output generation.

\textbf{Output generation.} The three parties cooperatively generate the output of the protocol as follows.

First, each party $P_i$ opens the commitment to its share $[c]_i$ of the cut-and-choose indicator, and reconstructs $c = (c^{(1)}, \cdots, c^{(s)})$.

Second,  for each run with $c^{(j)} = 1$, the commitments to the shares $T^j_{h(k+1)}$ $[n_{k-1}+1..n_k]$ and $T^j_{h(k+2)}[n_{k-1}+1..n_k]$ of each party $P_k$'s input (selected in the input selection circuit) are opened,  enabling $P_k$ to reconstruct this input and do consistence check by comparing the input with its random input. If the reconstructed input value does not equal to its random input value, $P_k$ aborts; otherwise, $P_k$ uncommits to $T^j_i[n_{k-1}+1..n_k]$. This forces the corrupted party to follow the protocol to select random inputs for computation verification.

Third, for each run with $c^{(j)} = 1$,  each party $P_k$ uncommits its internal state share $T^j_k[n + 1..n + q - m]$ and output share $T^j_k[n + q - m + 1..n + q]$ for $k \in \{1,2,3\}$. Then, each party $P_i$ ($i \in \{1,2,3\}$) ensembles $\{T^j_k[1..n+q]\}_{k=1}^3$, and computes all wire values $T^j[1..n+q] = \oplus_{k=1}^3 T^j_k[1..n+q]$ of the circuit for the $j$th run, and checks if the circuit computation is correct. If any error is found, $P_k$ aborts.

Finally,  for each run with $c^{(j)} = 0$, the commitments to the output shares $T^j_k[n + q - m + 1..n + q]$ for $k \in \{1,2,3\}$ are opened, and all the output values are reconstructed by each party. Every party then checks if all the output values are equal. If so, output the output value; otherwise, aborts.

Our detailed protocol is demonstrated in Fig. \ref{fig:3pcfromcc}.

\begin{figure}
\centering
\begin{boxedminipage}{\linewidth}
\begin{center}\textbf{Protocol} $\pi_{\mathtt{3PC}}^{\mathtt{mal}}(P_1, P_2, P_3)$\footnote{In this protocol, both AND and XOR operations are bitwise, if one or both operands are bit strings.}\end{center}
\textbf{Input:} $P_i$'s holds input $x^0_i \in \{0,1\}^{n_i - n_{i-1}}$ with $i \in \{1,2,3\}$ \\
\textbf{Auxiliary Inputs:} Statistics parameter $s$, a circuit $C = (n, m, q, F, S, G)$ representing the computed functionality
\begin{itemize}
\item[1. ]\textbf{Input preparation:} \\
For $i \in \{1,2,3\}$:
\begin{itemize}
\item[ - ] $P_i$ do input preparation (see Fig. \ref{fig:input-preparation}), and finally holds $([I_1]_i, [I_2]_i, [I_3]_i)$, $[c]_i$, and commitments to $[x^0_i]$, $x^1_i$, $[c]_i$
\end{itemize}

\item[2. ]\textbf{Circuit computation:}
\begin{itemize}
\item[ - ] Select proper inputs: \\
For $i \in \{1,2,3\}$: For $j \in \{1, \cdots, s\}$: For $k \in \{1,2,3\}$:
\begin{itemize}
\item[$\diamond$] $P_k$ computes $[t_{ij}]_k \leftarrow [x^{\sigma_{ij}}_i]_k \oplus [x^{1-\sigma_{ij}}_i]_k$
\item[$\diamond$] If $i = k$ then, $P_k$ computes $[c^{(j)}]_k \leftarrow [c^{(j)}]_k \oplus \sigma_{ij}$
\item[$\diamond$] Compute $[t_{ij} c^{(j)}]_k$ as follows:
\begin{itemize}
\item[$\circ$] $P_k$ generates $r_{i,j,k} \in_R \{0,1\}^s$
\item[$\circ$] $P_k$ sends $[t_{ij}]_k$, $[c^{(j)}]_k$ and $r_{i,j,k}$ to $P_{h(k+1)}$
\item[$\circ$] $P_k$ receives $[t_{ij}]_{h(k-1)}$, $[c^{(j)}]_{h(k-1)}$ and $r_{i,j,h(k-1)}$ from $P_{h(k-1)}$
\item[$\circ$] $P_k$ computes $[t_{ij} c^{(j)}]_k \leftarrow [t_{ij}]_k [c^{(j)}]_k \oplus [t_{ij}]_k [c^{(j)}]_{h(k-1)} \\ \oplus [t_{ij}]_{h(h-1)} [c^{(j)}]_k \oplus r_{i,j,k} \oplus r_{i,j,h(k-1)}$
\end{itemize}
\item[$\diamond$] $P_k$ computes $[x_{ij}]_k \leftarrow [t_{ij} c^{(j)}]_k \oplus [x^{\sigma_{ij}}_i]_k$
\end{itemize}

\item[ - ] Compute the circuit: \\
Now, parties $P_k$ ($k \in \{1,2,3\}$) hold $([x_{1j}]_k, [x_{2j}]_k,[x_{3j}]_k)$ ($j \in \{1, \cdots, s\}$) as inputs. Let $x_j = x_{1j} || x_{2j} || x_{3j}$, then $x_j \in \{0,1\}^n$. The input bits $\{x_j\}_s$ can be regarded as an $n \times s$ bit matrix. Letting $v_w \in \{0,1\}^s$ ($w \in \{1,\cdots,n\}$) denote the $w$th column of the matrix input to input wire $w$, the input can be written as $\{v_w\}_n$. The circuit with $s$ copies of inputs can be computed in parallel as follows.

For $k \in \{1,2,3\}$: For $j \in \{1, \cdots, s\}$: $T^j_k[1..n] \leftarrow [x_{1j}]_k || [x_{2j}]_k || [x_{3j}]_k$

For $g \in \{n + 1, \cdots, n + q\}$: $a \leftarrow F(g)$, $b \leftarrow S(g)$
\begin{itemize}
\item[$\diamond$] If $g$ is XOR gate: For $i \in \{1,2,3\}$: $[\delta^s(g)]_i \leftarrow [\delta^s(a)]_i \oplus [\delta^s(b)]_i$
\item[$\diamond$] If $g$ is AND gate (see Fig. \ref{fig:and-com}): For $i \in \{1,2,3\}$:
\begin{itemize}
\item[ $\circ$ ] $P_i$ generates $r_i \in_R \{0,1\}^s$
\item[ $\circ$ ] $P_i$ sends $[\delta^s(a)]_i$, $[\delta^s(b)]_i$ and $r_i$ to $P_{h(i+1)}$
\item[ $\circ$ ] $P_i$ receives $[\delta^s(a)]_{h(i-1)}$, $[\delta^s(b)]_{h(i-1)}$ and $r_{h(i-1)}$ from $P_{h(i-1)}$
\item[ $\circ$ ] $P_i$ computes $[\delta^s(g)]_i \leftarrow [\delta^s(a)]_i [\delta^s(b)]_i \oplus [\delta^s(a)]_i [\delta^s(b)]_{h(i-1)} \\ \oplus [\delta^s(a)]_{h(i-1)} [\delta^s(b)]_i \oplus r_i \oplus r_{h(i-1)}$
\end{itemize}
\item[$\diamond$] Let $\delta^s(g) = (\delta_1(g), \cdots, \delta_s(g))$. For $k \in \{1,2,3\}$: For $j \in \{1, \cdots, s\}$: $T^j_k[g] \leftarrow [\delta_j(g)]_k$
\end{itemize}

\end{itemize}

\item[3. ]\textbf{Transcript commitment:} \\
For $k \in \{1,2,3\}$: For $j \in \{1, \cdots, s\}$:
\begin{itemize}
\item[ - ] $P_k$ commits to $T^j_k[1..n]$ corresponding to input wires
\item[ - ] $P_k$ commits to $T^j_k[n + 1..n + q - m]$ corresponding to internal gate outputs
\item[ - ] $P_k$ commits to $T^j_k[n + q - m + 1..n + q]$ corresponding to output wires
\end{itemize}

\end{itemize}
\end{boxedminipage}
\end{figure}

\begin{figure}[!t]
\begin{boxedminipage}{\linewidth}
\begin{itemize}
\item[4. ]\textbf{Output generation:}
\begin{itemize}
\item[ - ] Construct cut-and-choose indicator: each party $P_i$ uncommits shares $[c]_i$ for $i \in \{1,2,3\}$, and gets:
$c = (c^{(1)}, \cdots, c^{(s)})$

\item[ - ] Check random inputs: For $j \in \{1,\cdots,s\}$: If $c^{(j)}=1$ then: For $k \in \{1,2,3\}$:
\begin{itemize}
\item[$\diamond$] $P_{h(k+1)}$ uncommits $T^j_{h(k+1)}[n_{k-1}+1 .. n_k]$
\item[$\diamond$] $P_{h(k+2)}$ uncommits $T^j_{h(k+2)}[n_{k-1}+1 .. n_k]$
\item[$\diamond$] $P_k$ checks whether \\
$x^j_k = T^j_1[n_{k-1}+1 .. n_k] \oplus T^j_2[n_{k-1}+1 .. n_k] \oplus T^j_3[n_{k-1}+1 .. n_k]$\\
If not so, $P_k$ aborts; Otherwise, $P_k$ uncommits $T^j_k[n_{k-1}+1 .. n_k]$
\end{itemize}
\item[ - ] Check circuit computation: For $j \in \{1,\cdots,s\}$: If $c^{(j)}=1$ then: \\
For $k \in \{1,2,3\}$:
$P_k$ uncommits $T^j_k[n + 1..n + q - m]$ and $T^j_k[n + q - m + 1..n + q]$\\
For $k \in \{1,2,3\}$:
\begin{itemize}
\item[$\diamond$] $P_k$ ensembles $T^j_i[1..n+q]$ for $i \in \{1,2,3\}$
\item[$\diamond$] $P_k$ computes $T^j[1..n+q] = \oplus_{k=1}^3 T^j_k[1..n+q]$, and checks if the circuit computation is correct. If any error is found, $P_k$ aborts.
\end{itemize}
\item[ - ] Construct outputs:
\begin{itemize}
\item[$\diamond$] For $j \in \{1,\cdots,s\}$: If $c^{(j)}=0$ then: For $k \in \{1,2,3\}$: \\
$P_k$ uncommits $T^j_k[n + q - m + 1..n + q]$
\item[$\diamond$] For $j \in \{1,\cdots,s\}$: If $c^{(j)}=0$ then: For $k \in \{1,2,3\}$: \\
$P_k$ computes $y_j = \oplus_{i=1}^3 T^j_i[n + q - m + 1..n + q]$
\item[$\diamond$] For $k \in \{1,2,3\}$: $P_k$ checks if all $y_j$ with $c^{(j)}=0$ are equal. If not so, aborts; otherwise, output any $y_j$ with $c^{(j)}=0$ as the common output.
\end{itemize}

\end{itemize}
\end{itemize}
\end{boxedminipage}
\caption{Our 3PC protocol from cut-and-choose in the presence of a malicious adversary corrupting a party}
\label{fig:3pcfromcc}
\end{figure}

\subsection{Security Analysis}\label{sec:analysis}

In this section, we first show that the three-party commitment scheme in Fig. \ref{fig:commit} is perfectly binding and perfectly hiding, and then show that our protocol is information-theoretically secure against one malicious corruption.

Theorem 2 states the security of the commitment shceme $\pi_{\mathtt{com}}(P_1, P_2, P_3)$.\\\\
\textbf{Theorem 2.} \emph{Protocol $\pi_{\mathtt{com}}(P_1, P_2, P_3)$ is of perfect binding and perfect hiding in the presence of a malicious adversary corrupting at most one of the parties.}  \\\\
\textbf{Proof.} Without loss of generality, we assume that $P_1$ is the committer and $P_2$, $P_3$ are verifiers. Then, we consider the cases where $P_1$ is corrupted and where $P_2$ is corrupted. (the cases of $P_2$ and $P_3$ are essentially symmetric).

If $P_1$ is corrupted, the honest verifiers $P_2$ and $P_3$ will hold two shares of the committed data $d$, which will uniquely determine $d$ due to Shamir's secret sharing, and $P_1$ cannot open the commitment to any other value except the original value committed in the uncommitment phase. Thus, perfect binding holds.

If $P_2$ is corrupted, it receives a share of $d$ in the commitment phase, which is a uniformly random number in $\mathbb{F}_p$ and carries no information about $d$. Thus the perfect hiding holds. $\Box$\\

The Theorem 3 states the security of our 3PC protocol with malicious security.\\\\
\textbf{Theorem 3.} \emph{Protocol $\pi_{\mathtt{3PC}}^{\mathtt{mal}}(P_1, P_2, P_3)$ is secure in the presence of a malicious adversary corrupting at most one of the parties, and achieves a cheating probability $2^{-s}$ where $s$ is the statistical parameter.}  \\\\
\textbf{Proof.} Given that the commitment scheme $\pi_{\mathtt{com}}(P_1, P_2, P_3)$ is of perfect binding and perfect hiding, it is obvious that all the messages caused by commitments can be perfectly simulated, and the commitment data cannot be changed once it is committed. Thus, in the simulation following, we omit the simulation of messages caused by commitments, and use the three-party commitment protocol as an ideal functionality.

Since the three parties are symmetric in the protocol, we consider only the case where $P_1$ is corrupted. We show that the real and simulated interactions are indistinguishable to all environments, in every part of the protocol. Recall that the view of an environment consists of messages sent from honest parties to the corrupted party, as well as the final outputs of the parties.
\begin{itemize}
\item \textbf{Input sharing.} In this phase, $P_1$  receives $[I_2]_1$ from honest $P_2$, and $[I_3]_1$ from honest $P_3$. This can be simulated with two uniformly random bit string of the same bit lengths as the two received messages. Specifically, message $[I_2]_1 = (([x^{\sigma_{2,1}}_2]_1, [x^{1-\sigma_{2,1}}_2]_1), \cdots, ([x^{\sigma_{2,s}}_2]_1, [x^{1-\sigma_{2,s}}_2]_1))$ can be simulated with a random string $t_1 \in \{0,1\}^{s|x^0_2|}$, while message $[I_3]_1 = (([x^{\sigma_{3,1}}_3]_1, [x^{1-\sigma_{3,1}}_3]_1),$ $ \cdots, ([x^{\sigma_{3,s}}_3]_1, [x^{1-\sigma_{3,s}}_3]_1))$ can be simulated with a random string $t_2 \in \{0,1\}^{s|x^0_3|}$. Due to the XOR secret sharing used, the real and simulated views are identically distributed.
\item\textbf{Circuit computation.} This phase comprises two steps:
\begin{itemize}
\item \emph{Select proper inputs}. \\ $P_1$ receives $\{([t_{1,j}]_3, [t_{2,j}]_3, [t_{3,j}]_3)\}_{j=1}^s$, $[c]_3$ and $\{(r_{1,j,3}, r_{2,j,3}, r_{3,j,3})\}_{j=1}^{s}$ \\ where $[t_{1,j}]_3 || [t_{2,j}]_3 || [t_{3,j}]_3, r_{1,j,3} || r_{2,j,3} || r_{3,j,3} \in \{0,1\}^n$, $[c]_3 \in \{0,1\}^s$ and $[t_{i,j}]_3 = [x^{\sigma_{i,j}}_i]_3 \oplus [x^{1 - \sigma_{i,j}}_i]_3$ for $i \in \{1,2,3\}$ and $j \in \{1, \cdots, s\}$. These messages are all uniformly random strings for $P_1$, except that $[x_1^{\sigma_{1,j}}]_3$ and $[x_1^{1-\sigma_{1,j}}]_3$ had been generated by $P_1$ and thus $[t_{1,j}]_3$ is known exactly to $P_1$. Therefore, these messages can be simulated with strings $t_1 \in_R \{0,1\}^{ns}$, $t_2 \in_R \{0,1\}^s$ and $t_3 \in_R \{0,1\}^{ns}$, respectively, with a restraint that substrings $[t_{1,j}]_3$ are fixed in the correponding places of $t_1$. This way, the real and simulated views of all environments are identical.
\item \emph{Compute the circuit}. \\
This can be simulated gate by gate until the computation is completed. Given a current gate, it is simulated in terms of its operation as follows.
\begin{itemize}
\item \textit{XOR}: Party $P_i$ sends or receives nothing, so there
is nothing to be simulated.
\item \textit{AND}: Let $a, b \in \{0,1\}^s$ denote the two input bit strings to the gate. Let $r_i, r_{h(i-1)} \in \{0,1\}^s$ be the two uniformly random bit strings generated by $P_i$ and $P_{h(i-1)}$ for the AND computation, respectively. During the gate computation, $P_i$ receives $[a]_{h(i-1)}$, $[b]_{h(i-1)}$ and $r_{h(i-1)}$ from $P_{h(i-1)}$. From the circuit computation of the protocol, we can see that $\{[a]_k\}_{k=1}^3$ and $\{[b]_k\}_{k=1}^3$ are all independent random variables subject to $a = \oplus_{k=1}^3 [a]_k$ and $b = \oplus_{k=1}^3 [b]_k$. Thus, the three messages can again be simulated by three uniformly random bit strings. Both the real and simulated views are also identical.
\end{itemize}
\end{itemize}
Thus, the entire phase of circuit computation can be perfectly simulated.
\item\textbf{Transcript commitment.} This phase can be perfectly simulated due to the perfect hiding of the commitment scheme.
\item\textbf{Output generation.}  This phase make sure that any deviation from the protocol can be detected except with negligible probability in statistical parameter $s$, making use of the perfect binding of the commitment scheme used.
\begin{itemize}
\item Construct cut-and-choose indicator: The protocol either open the shares correctly, and thus all parties can construct $c$, or is aborted for some cheating detected. Namely, once all the shares of $c$ is committed, they cannot be modifed in all runs of computation.
\item Check random inputs: The protocol either properly selects the random inputs as inputs to runs for computation verification, or is aborted for some cheating detected. This makes sure that all runs for computation verification should use random inputs, otherwise the protocol will be aborted and noting is revealed.
\item Verify circuit computation: The protocol either successfully verify the correctness of all the runs for computation verification, or is aborted for some cheating detected. This makes sure all runs for computation verification are correctly performed.
\item Construct output: The protocol either correctly constructs the outputs, or is aborted for some cheating detected. This makes sure that all runs for output computation are consistently performed.
\end{itemize}
\end{itemize}

From the above, we can see that for semi-honest adversaries, the protocol would be carried out successfully; for malicious adversaries, any cheating would be caught except with a negligible probability and the protocol would be aborted. In a word, all environments' real and simulated views are indistinguishable.

To cheat successfully, the adversary has to guess the value of $c$, thus the cheating probability is $2^{-s}$.  $\Box$

\subsection{Efficiency Discussion}\label{sec:efficiency}

Now we discuss the efficiency of the proposed protocol. The computation complexity is mild since there are no cryptographic operations, and most of the operations are bitwise XOR and AND computations. The communication complexity is reduced compared to the counterparts based on verifiable secret sharing. Specifically, the communication round amounts to the depth of the circuit plus a constant number, while that of the protocols based on verifiable secret sharing may be several times of the circuit depth.

\section{Practical Considerations of Our Protocol}\label{sec:prac-consid}

In practical scenarios, malicious security may be much stronger than actual security demand. For example, if a protocol ensuring that $99.5\%$ of all cheatings are caught is sufficient for practical applications, the a covert security level fixing $s = 8$ can be used, and the protocol is converted into covertly secure one with deterrent $1-2^{-8}$. In other words, we can fix $s$ properly at a small value according to practical requirements, and get a protocol with appropriate deterrent.

Furthermore, at most of the time in practice, computational security is enough. Thus, the unifromly random bits can be generated by a pseudo-random generator, and the commitments to the transcripts of $s$ runs can be replaced with the commitments to hash values of the transcripts, degrading the original information-theoretically secure protocol to a computationally secure one, but with a better efficiency.

\section{Conclusion}\label{sec:conclusion}

In this work, we apply cut-and-choose idea in the information-theoretic setting to construct a maliciously secure protocol for three-party computations (3PC) with one corruption. This yields an efficient and information-theoretically secure 3PC protocol for Boolean circuit computation. Asymptotically, we achieve a cheating probability of $2^{-s}$ where $s$ is the number of runs of circuit computation. This is similar to the setting where cut-and-choose technique is used based on Yao's garbled circuits \cite{lindell2013fast}.


\bibliographystyle{alpha}
\bibliography{reference}

\end{document}